\documentclass[sigconf,screen]{acmart}
\AtBeginDocument{%
  \providecommand\BibTeX{{%
    \normalfont B\kern-0.5em{\scshape i\kern-0.25em b}\kern-0.8em\TeX}}}

\usepackage{listings}
\usepackage{tabularx}
\usepackage{adjustbox}
\usepackage{geometry}
\usepackage{makecell}
\usepackage{multirow}
\usepackage{enumitem}

\setcopyright{acmcopyright}
\copyrightyear{2018}
\acmYear{2018}
\acmDOI{XXXXXXX.XXXXXXX}

\acmConference[ITiCSE 2024]{Innovation and Technology in Computer Science Education }{July 08--10,
  2024}{Milan, Italy}
\acmPrice{15.00}
\acmISBN{978-1-4503-XXXX-X/18/06}

\title[Can ChatGPT Play the Role of a TA in an Introductory Programming Course?]{Can ChatGPT Play the Role of a Teaching Assistant in an Introductory Programming Course?}

\author{Anishka}
\email{anishka20282@iiitd.ac.in}
\affiliation{%
  \institution{IIIT Delhi}
  \state{Delhi}
  \country{India}
}
\author{Atharva Mehta}
\email{atharva20038@iiitd.ac.in}
\affiliation{%
  \institution{IIIT Delhi}
  \state{Delhi}
  \country{India}
}
\author{Nipun Gupta}
\email{nipun20089@iiitd.ac.in}
\affiliation{%
  \institution{IIIT Delhi}
  \state{Delhi}
  \country{India}
}
\author{Aarav Balachandran}
\email{aarav21302@iiitd.ac.in}
\affiliation{%
  \institution{IIIT Delhi}
  \state{Delhi}
  \country{India}
}

\author{Dhruv Kumar}
\email{dhruv.kumar@iiitd.ac.in}
\affiliation{%
  \institution{IIIT Delhi}
  \state{Delhi}
  \country{India}
}
\author{Pankaj Jalote}
\email{jalote@iiitd.ac.in}
\affiliation{%
  \institution{IIIT Delhi}
  \state{Delhi}
  \country{India}
}

\begin{document}



\renewcommand{\shortauthors}{Anony Mous et al}

\begin{abstract}
The emergence of Large language models (LLMs) is expected to have a major impact on education. This paper explores the potential of using ChatGPT, an LLM, as a virtual Teaching Assistant (TA) in an Introductory Programming Course. We evaluate ChatGPT's capabilities by comparing its performance with that of human TAs in some of the important TA functions. The TA functions which we focus on include (1) grading student code submissions, and (2) providing feedback to undergraduate students in an introductory programming course. Firstly, we assess ChatGPT's proficiency in grading student code submissions using a given grading rubric and compare its performance with the grades assigned by human TAs. Secondly, we analyze the quality and relevance of the feedback provided by ChatGPT. This evaluation considers how well ChatGPT addresses mistakes and offers suggestions for improvement in student solutions from both code correctness and code quality perspectives. We conclude with a discussion on the implications of integrating ChatGPT into computing education for automated grading, personalized learning experiences, and instructional support.
\end{abstract}




\received{20 February 2007}
\received[revised]{12 March 2009}
\received[accepted]{5 June 2009}

\maketitle

\section{Introduction}
ChatGPT, a large language model (LLM) developed by OpenAI, represents a groundbreaking advancement in natural language processing and artificial intelligence \cite{openai_introducing_2022}. ChatGPT demonstrates an exceptional capability to understand and generate human-like text responses, enabling seamless user interactions. Trained on extensive datasets, ChatGPT can engage in a wide range of conversations, making it a versatile tool for answering questions, providing explanations, and even assisting with creative writing. Numerous research studies have been conducted in the computing education community concerning Large Language Models (LLMs). Educators have assessed the performance of these LLMs in generating precise solutions to questions posed to students both within and beyond the classroom setting \cite{Denny2023CopilotCS1, Finnie-Ansley2022CS1, wermelinger2023Copilot, Reeves2023Parsons, finnie-ansley2023CodexCS2, Savelka2023MCQAndCode, Ouh2023Java, Cipriano2023GPT-3OOP, Daun2023Software, Malinka2023Security}. Additionally, some studies have explored the application of LLMs in generating fresh programming exercises, along with corresponding code solutions and explanations \cite{sarsa2022AutoGenerate, wermelinger2023Copilot, Leinonen2023CodeExplanation}. Notably, certain investigations have delved into the utility of ChatGPT for code debugging, providing suggestive fixes to code errors \cite{Leinonen2023ExplainError, MacNeil2023CodeExplain}. Furthermore, the generation of personalized feedback for students based on their code submissions has also been a subject of inquiry in a research study conducted by Balse et al \cite{Balse2023Feedback}.

In this work, we analyze the possibility of using ChatGPT as a virtual Teaching Assistant (TA) in an introductory programming course. An introductory programming course often has large enrollments involving many programming assignments. For evaluating assignments and giving feedback to students, often a large number of TAs are needed, which also brings in issues like non-uniformity in grading or providing feedback. The quality of the course can improve tremendously if the TA functions can be partially automated and the quality of TA support improved - this is the main motivation behind this work. ChatGPT or any LLM in general offers an alternate approach to grading student code submissions, focusing on the quality of logic and understanding rather than just the end result. Unlike automated test case-based grading, LLMs can comprehend partially correct logic in student submissions, even if they don't pass the test cases. This method ensures that students are rewarded for their understanding and problem-solving skills, not just for achieving the correct output, fostering a deeper and encouraging learning experience. We specifically look at the following TA functions: (1) grading student code submissions, and (2) providing qualitative feedback to undergraduate computer science students in an introductory programming course (equivalent to CS1 programming). 

\noindent We begin with comparing ChatGPT's proficiency in grading student code with the grades assigned by human TAs with respect to code functionality. The human TAs did not grade the student code with respect to code quality in the dataset we had access to. Hence, we grade the student code with respect to code quality using ChatGPT and compare it with the respective code quality metrics (generated using a library) to understand if ChatGPT correctly analyses the code quality. Finally, we conduct an assessment of the quality and relevance of the feedback provided by ChatGPT. This evaluation considers its effectiveness in addressing mistakes and offering improvement suggestions in student solutions, taking into account both code correctness \cite{Balse2023Feedback} and code quality perspectives.

\noindent Our paper concludes by discussing the potential implications of integrating ChatGPT into computing education. The envisioned outcomes encompass automated grading, personalized learning experiences, and instructional support.


\noindent This paper finds answers to the following research questions:
\begin{itemize}[leftmargin=*]
    \item \textbf{RQ1:} In terms of grading the student code submissions, how does ChatGPT's evaluation compare with the evaluation done by human TAs?
    \item \textbf{RQ2:} How effective is ChatGPT in giving feedback to students in terms of identifying mistakes in their code and possible code improvements?
\end{itemize}

\section{Related Work}

Within the computing education research community, there have been several studies and reports on LLMs \cite{becker2023ProsAndCons, Malinka2023Security, Daun2023Software, Denny2023CopilotCS1, Finnie-Ansley2022CS1, wermelinger2023Copilot, Savelka2023MCQAndCode, Reeves2023Parsons, finnie-ansley2023CodexCS2, Ouh2023Java, Cipriano2023GPT-3OOP, sarsa2022AutoGenerate, Leinonen2023CodeExplanation, Leinonen2023ExplainError, MacNeil2023CodeExplain, Balse2023Feedback, joshi2023chatgpt}. Becker et al. \cite{becker2023ProsAndCons} examine the various challenges and opportunities associated with the utilization of AI code generation tools such as OpenAI Codex \cite{codex}, DeepMind AlphaCode \cite{li_competition-level_2022}, and Amazon CodeWhisperer \cite{amazon_codewhisperer}. Instructors can use these AI tools to generate new content and exercises for the students as well as generate simplified explanations of any computer science concept. Students can use these tools for getting boilerplate code to get started on their programming assignments and also for debugging errors encountered during programming. At the same time, students can misuse these tools to get complete answers to graded assignments and exams. Similar challenges and opportunities have also been discussed in other studies \cite{Denny2023CopilotCS1, Daun2023Software}. A good number of research studies have specifically analyzed the accuracy of LLMs, such as OpenAI Codex, GPT-3, ChatGPT (GPT-3.5 and GPT-4), in generating solutions for programming assignments across various computer science courses \cite{Denny2023CopilotCS1, Finnie-Ansley2022CS1, wermelinger2023Copilot, Savelka2023MCQAndCode, Reeves2023Parsons, finnie-ansley2023CodexCS2, Savelka2023MCQAndCode, Ouh2023Java, Cipriano2023GPT-3OOP, Daun2023Software, Malinka2023Security}. These studies show that the performance of LLMs is comparable to and sometimes even better than the performance of an average student in the class.


Leinonen et al. \cite{Leinonen2023ExplainError} conducted analyses to evaluate OpenAI Codex's capability in explaining different error messages encountered during code execution and the effectiveness of the corresponding code fixes suggested by Codex. This functionality is particularly valuable for debugging programs. Balse et al. \cite{Balse2023Feedback} explore the potential of GPT-3 in delivering detailed and personalized feedback for programming assessments, addressing a challenging task in large classes of students. Additionally, Sarsa et al. \cite{sarsa2022AutoGenerate} discuss the automatic generation of programming exercises and solutions using OpenAI Codex, offering instructors a valuable resource to create innovative programming assignments for their students.

Additionally, several studies analyze the LLM models' ability to generate code explanations, contrasting the quality of these explanations with those provided by students \cite{sarsa2022AutoGenerate, wermelinger2023Copilot, Leinonen2023CodeExplanation, MacNeil2023CodeExplain}. Some studies have also evaluated the perception and current usage of LLMs among instructors and students \cite{Lau2023Instructor, budhiraja2024Jarvis, joshi2023interviews}. 

Recent research has also explored the application of these LLMs in roles traditionally held by teaching assistants (TAs), particularly in programming courses \cite{FinnieRobots, Hellas_2023, Nilsson1779778}. These studies focus on functions like answering student questions \cite{hicke2023aita} and assisting in problem-solving by offering hints \cite{lee2023learning}. However, this research is still in very early stages. The existing work is far from evaluating the full range of responsibilities typically handled by a TA. As part of our work, we aim to evaluate some of the TA responsibilities such as grading student work and offering constructive feedback for improvement.

Existing approaches to automated grading of programming assignments involve evaluating student submissions by comparing them to predefined test cases and identifying coding errors \cite{fordentA-bot}. Additionally, certain automated grading systems have incorporated the use of instructor-provided reference solutions, assessing student work by comparing it against the logical structure of these model codes using semantic analysis \cite{fanCS1-AutomatedGrading, liu_autograde_assignments}. However, these tools have several limitations. For instance, test case-based grading can miss logical correctness in student code. On the other hand, reference code-based auto-graders may not account for the variety of correct solutions. There are often multiple valid approaches to solving a programming problem, and not all these approaches may align with the reference code provided. In contrast, our work aims to explore the application of LLMs in automated grading and qualitative feedback for overcoming the aforementioned limitations in the existing approaches.
\section{Methodology and Experimental Design}
We have used an experimental research design to determine the viability of ChatGPT as a teaching assistant (TA) for the introduction to programming course (CS1) offered to first-year undergraduate students at a premier engineering institute in India. In this section, we detail a series of experiments designed to evaluate ChatGPT's capabilities in grading student code submissions and providing relevant feedback to students. We utilized OpenAI's GPT 3.5 model, specifically the '\textit{text-davinci-003}' version, for our experiments. To facilitate large-scale experimentation, we opted for the API version of the AI model rather than the web interface. In this section, we describe the experiments we performed. In the next section, we will show the analysis of the data from the experiments. 
\vspace{-1pt}
\subsection{Dataset}
In our assessment, we focused on three take-home programming assignments from the CS1 course. We opted to use our own dataset rather than a public one, as it included the grades provided by TAs for each assignment. This allowed us to evaluate the effectiveness of LLM for partial marking based on functionality in the scoring process. Python \cite{python} was the specific programming language used to teach the CS1 course. There were a total of 6783 student submissions across the three assignments. Each assignment consisted of multiple questions of varying difficulties. One student code submission refers to code submission for one question in the assignment. Hence, there could be multiple code submissions from the same student for different questions in the assignment. Table \ref{table:assignment_details} contains more details about the number and type of questions and total student submissions for each assignment.
\vspace{1pt}
\begin{table}[htbp]
	\small
	\vspace{-1em}
	\begin{tabular}{|p{2cm}|p{3cm}|p{1.8cm}| } 
		\hline
		\textbf{Assignment No. (Number Of Questions)} & \textbf{Question Type} & \textbf{Total Student Code Submissions}\\
		\hline
            1 (10) & Basic Algorithms, Lists, Variables and Basic Arithmetic  & 2795 \\
            \hline
            2 (8) & Dictionaries, Lists and Tuples.  & 2442 \\
            \hline
            3 (5) & Nested dictionaries, Advanced data structures and classes  & 1546 \\
            \hline
	\end{tabular}
 \vspace{1em}
	\caption{\textbf{Assignment Details}}
        \vspace{-1em}
	\label{table:assignment_details}
\end{table}



\noindent 
The CS1 course consisted of 32 highly proficient TAs which included B.Tech students from junior and senior years as well as M.Tech students. These TAs were selected based on their exceptional performance in this course when they were enrolled in it in the past offerings of the course, ensuring a high standard of grading and understanding of the course material. Each TA was responsible for a group of 15-20 students. Their duties included conducting tutorials, addressing student queries, and grading the assignments of the students assigned to them. In contrast to many existing online and offline platforms that grade code submissions strictly based on input/output correctness, TAs also consider partial functionality in their evaluations. This approach allows students to receive partial credit for their submissions, recognizing the correctness of certain logical elements even if the overall code doesn't pass all test cases. This method acknowledges the importance of code efficiency, readability, and adherence to coding standards, ensuring that students are rewarded for their partial successes and accurate logic implementation, not just the final output generated by the code. We are exploring the feasibility of utilizing ChatGPT for grading purposes precisely because of this aspect of awarding partial credits to the student code submissions.

\subsection{Experiment 1: Grading Student Code Submissions}

In Experiment 1, we assess ChatGPT's capability to grade student code submissions, focusing on both correctness and quality. The experiment involves providing ChatGPT with the assignment question, solution rubric, and student code from an Introduction to Programming course. ChatGPT is tasked with grading the code based on functionality and quality, where functionality scores are assigned as 2 for completely correct, 1 for partially correct, and 0 for incorrect or empty submissions. For quality assessment, ChatGPT assigns grades based on Halstead metrics and code modularity. 
\noindent A sample prompt provided to ChatGPT is given below:\newline\newline
\noindent \textit{"""\textbf{Context}: You are a teaching assistant for an Introduction To Programming Course, and your task is to grade the student code submission using the provided rubric. The code is written in Python programming language.}

\noindent\textit{\textbf{Question}: Integration through computation. Suppose the velocity of a rocket at a time t is given by: f(t) = 2000ln[140000140000 - 2100t] - 9.8t Take as input starting time (a) and ending time (b), and find the distance covered by a rocket between time a and b. For computationally determining the distance, work with time increments (delta) of 0.25 seconds. You can use the math module of Python for this problem. (Hint: Compute the velocity at time t and t+delta, take the average, and then compute the distance traveled in this delta time duration. Start from a and keep computing in increments of delta till b.)}

\noindent\newline\textit{Given this question you are supposed to mark the student code based on functionality. If the code is completely correct give 2 marks. If the code is partially correct give 1 mark. If the code is incorrect give 0 marks. If the code is empty give 0 marks. Do not return the distribution of marks. Just return the final marks. In addition to the functionality score also give a quality score on a scale of 1-5 based on Halstead metrics and code modularity. Do not correct the previous solution or solve the question just give the final score as output without any explanation or extra text. Follow all the instructions carefully.}\newline
\noindent \textit{\textbf{Student Code : \{Student Code Goes Here\}}}\newline
\textit{The output should be of the format : \newline
Functionality Score : \newline
Halstead Metrics-based Quality Score : \newline
Modularity-based Quality Score : \newline
Overall Quality Score : """}\newline

\noindent We evaluate ChatGPT's proficiency in grading student code submissions in terms of code correctness and code quality in the following manner: \newline
\noindent\textbf{(1) Code Correctness: } We compare the scores awarded by ChatGPT with those given by teaching assistants (TAs). We compute the absolute deviation between ChatGPT's assigned scores and the TA-assigned scores for each student and the average and maximum absolute deviations for each assignment.
\noindent We also compute the correlation between the ChatGPT-assigned score and TA-assigned score using Pearson Correlation Coefficient \cite{wiki:Pearson_correlation_coefficient}.\newline


\noindent\textbf{(2) Code Quality: } Since the TAs did not grade the student submissions based on code quality,  We assess the quality of a code through two metrics computed using the Radon Python package \cite{python_radon}: Halstead's effort score and modularity. 

\begin{itemize}[leftmargin=*]
    \item\textbf{Halstead Effort score:} Maurice Halstead proposed a set of measures to evaluate the complexity of a software program \cite{halstead1977metrics}. These metrics are based on the number of distinct operators and operands  and the total number of operators and operands in the program. From these, it computes the volume and difficulty measures. Effort score is the product of volume and difficulty and can be used to estimate the effort required to develop a program - the higher the effort score of a code, more effort is required.
    \item \textbf{Modularity:} This refers to how well the components of a program are separated and independent, making it easier to maintain and scale the program. We consider the number of functions present as our metric to assess the modularity of code. It is a good practice to divide the code into functions as it increases readability, and it is easier to debug a code divided into smaller logical sections. 
\end{itemize}

\noindent Utilizing ChatGPT for code quality evaluation, instead of automated tools like Radon, offers significant benefits. ChatGPT enhances the interpretability of complex quality metrics such as Halstead effort and modularity, compiling these complex quality metrics into a single qualitative score based on its understanding of the code.
This helps in combining these complex quality metrics into a single qualitative score for analysing the relative code quality. Additionally, this approach tests ChatGPT's ability to understand and interpret code quality with respect to Halstead and modularity metrics.

For comparison, we computed the ratio of the metrics obtained for each question from Radon package to the value obtained from ChatGPT. We compute the correlation between the code quality scores assigned by ChatGPT and the code quality metrics assigned by Radon to determine its proficiency in grading student code submissions based on code quality.

In this preliminary study, we focused mainly on Halstead metrics and modularity, while excluding other quality metrics such as logical lines of code (LLOC), comment analysis and cyclomatic complexity. 
In our future work, we aim to broaden our scope to include a wider array of metrics including the above mentioned metrics, to provide a more comprehensive assessment of code quality.

\subsection{\textbf{Experiment 2: Providing Feedback on Student Code Submissions}}
In this study, we evaluate the potential of ChatGPT to provide pertinent feedback and suggestions on student submissions, which is crucial for enhancing the learning experience of novice programmers, such as those enrolled in a CS1 course. We randomly select approximately 10 submissions from each assignment while ensuring each question is represented at least once. We then prompted ChatGPT to generate feedback for these code submissions and propose specific improvement changes. Here is a sample prompt provided to ChatGPT:\newline\newline
\noindent \textit{"""For the given code snippet, give feedback on code quality and suggest ideas for improvement:\newline
\textbf{Student Code: \{Student Code Goes Here\}}}
\textit{"""}\newline\newline
In order to evaluate the feedbacks generated by ChatGPT we use the following two criteria: 

\subsubsection{\textbf{Subjective Assessment: }}
This experiment is designed to evaluate the  capability of ChatGPT to provide constructive feedback on code snippets. We used a survey-based evaluation process in which we asked four B.Tech students to review the feedback generated by ChatGPT. The selected students were a combination of previous TAs and high performers in the course. They also had a strong background in programming. Each of these four students were provided with a form containing all the pairs of code snippet and corresponding feedback generated by ChatGPT. For each pair, the students had to classify ChatGPT's feedback into one of three predefined categories: 
\begin{itemize}[leftmargin=*]
\item\textbf{Minor Enhancement Feedback} 
\newline This category is for feedback where ChatGPT suggests subtle improvements to already functional code. The focus here is on enhancing aspects like readability, maintainability, and adherence to best practices, without fundamentally altering the code's core logic or structure. For instance, suggestions for refining code style as per PEP 8 guidelines in Python would fall under this category.
\item \textbf{Critical Improvement Feedback} 
\newline Feedback classified here is more substantial, targeting significant improvements or rectification of major issues. This might include suggestions for optimizing performance through notable changes (like altering algorithms or data structures), enhancing security measures, improving error handling, or making the code more scalable and maintainable. Such feedback often necessitates considerable adjustments to the code's structure or underlying logic.
\item \textbf{Erroneous Feedback}
\newline This category encapsulates feedback where ChatGPT's suggestions are incorrect, based on misunderstandings of the code's purpose, or are contextually inapplicable. It includes feedback that, if implemented, could potentially degrade the code's efficiency, security, or robustness. For example, incorrect solutions or recommendations that deviate from established best practices are classified here.

\end{itemize}
This subjective approach allows for an in-depth understanding of how students perceive the relevance and accuracy of the feedback provided by ChatGPT.

\subsubsection{\textbf{Objective Assessment:}}
 Alongside the above mentioned survey, a quantitative assessment was also conducted. In this phase, all feedback provided by ChatGPT was implemented into the respective code snippets. Subsequently, code quality metrics that were calculated in Experiment 1 were recalculated for these modified snippets and compared with the original code quality metrics. We also regraded the modified code snippets with respect to code correctness. This quantitative approach was crucial for objectively measuring the impact of ChatGPT's feedback on the overall code quality and correctness, providing a balanced evaluation alongside the qualitative survey evaluation.

\section{Evaluation \& Results}
\subsection{\textbf{Experiment 1: Grading Student Code Submissions}}
In this experiment, we assess how good a job ChatGPT does in assessing students' code for correctness and quality. Table \ref{table:code_correctness_grading} presents the mean deviation, max deviation, and correlation coefficient between ChatGPT-assigned and TA-assigned scores for code correctness. The overall value for each assignment is determined by aggregating the scores from all questions for each student and then calculating the mean and maximum deviation for these aggregated scores. Unfortunately, we could not conduct a question-wise analysis for Assignment-1 due to the unavailability of question-specific scores from the TAs for this assignment.\newline
High values of mean and max deviations in Table \ref{table:code_correctness_grading} suggest that there is a significant difference between TA-assigned scores and ChatGPT-assigned scores for all three assignments. At the same time, the correlation coefficient values indicate a moderately positive correlation between the TA-assigned scores and ChatGPT-assigned scores for assignment 2 and assignment 3, but the same cannot be said for assignment 1.

\begin{table}[htbp]
\small
\centering
\begin{tabular}{|c|c|c|c|c|c|}
\hline
\textbf{\shortstack{Assign \\ No.}} & \textbf{\shortstack{Q \\No.}} & \textbf{\shortstack{\\Mean\\ Deviation}} & \textbf{\shortstack{\\Max \\Deviation}} & \textbf{\shortstack{\\Total\\ Score}} & \textbf{\shortstack{\\Correlation\\ Coefficient}} \\ 
\hline
1 & \textbf{Overall} & \textbf{5.35} & \textbf{16.0} & \textbf{20.0} & \textbf{0.143}\\
\hline
\multirow{8}{*}{2}  & Q1 & 0.46 & 2.0 & 2 & -0.137\\
\cline{2-6}
 & Q2 & 0.71 & 2.0 & 2 & 0.373\\
\cline{2-6}
 & Q3 & 0.42 & 2.0 & 2 & 0.468\\
\cline{2-6}
 & Q4 & 0.89 & 2.0 & 2 & 0.251\\
\cline{2-6}
 & Q5 & 0.74 & 2.0 & 2 & 0.335\\
\cline{2-6}
 & Q6 & 0.20 & 2.0 & 2 & 0.494\\
\cline{2-6}
 & Q7 & 0.78 & 2.0 & 2 & 0.56\\
\cline{2-6}
 & Q8 & 0.61 & 2.0 & 2 & 0.61\\
 \cline{2-6}
 & \textbf{Overall} & \textbf{5.47} & \textbf{13.0} & \textbf{16.0} & \textbf{0.462}\\
\hline
\multirow{5}{*}{3} & Q1 & 1.08 & 2.0 & 2 & 0.01\\
\cline{2-6}
 & Q2 & 0.90 & 2.0 & 2 & 0.46\\
\cline{2-6}
 & Q3 & 0.68 & 2.0 & 2 & 0.19\\
\cline{2-6}
 & Q4 & 0.86 & 2.0 & 2 & 0.28\\
\cline{2-6}
 & Q5 & 0.91 & 2.0 & 2 & 0.19\\
 \cline{2-6}
 & \textbf{Overall} & \textbf{3.94} & \textbf{8.0} & \textbf{10.0} & \textbf{0.59}\\
\hline
\end{tabular}
\caption{ChatGPT's grading compared with that of human TAs for code correctness.}
\vspace{-2em}
\label{table:code_correctness_grading}
\end{table}
\noindent Table \ref{table:code_quality_grading} presents the correlation between the code quality scores assigned by ChatGPT (based on Halstead effort score and modularity), and the corresponding effort scores and modularity scores computed using Radon python package. Table \ref{table:code_quality_grading} suggests that there is not much correlation between the ChatGPT-assigned score and actual code quality score as all correlation scores are in the range $[-0.1, 0.2]$. 

\begin{table}[htbp]
\centering
\vspace{0em}
\begin{tabular}{|p{2.8cm}|p{2.2cm}|p{2.2cm}|}
\hline
\textbf{\shortstack{\\Assignment No.}} & \textbf{\shortstack{\\Effort Score}} & \textbf{\shortstack{\\Modularity}}\\
\hline
Assignment-1 & -0.024 & -0.012 \\ \cline{1-3}
Assignment-2 & -0.04 & 0.054\\ \cline{1-3}
Assignment-3 & 0.16 & 0.19\\ \cline{1-3}
\hline
\end{tabular}
\caption{Correlation coefficient between ChatGPT-assigned scores for code quality and code quality metrics generated using Radon package}
\vspace{-2em}
\label{table:code_quality_grading}
\end{table}
\noindent Based on this data, we conclude that currently, it may not be a good idea to rely on ChatGPT for grading student submissions either from code correctness or code quality perspective. Further research needs to be conducted in order to assess the suitability of using ChatGPT as a reliable tool for grading student submissions.

\subsection{\textbf{Experiment 2: Providing Feedback on Student Code Submissions}}
In this experiment, we are assessing the quality of suggestions provided by ChatGPT for code improvement in student code.
\subsubsection{\textbf{Subjective Assessment: }}Table \ref{table:code_feedback} presents the \% of feedback comments marked in each category by the expert programmers, and the average number of feedback comments generated per question.  
\begin{table}[htbp]
    \centering
    \begin{adjustbox}{max width=\textwidth}
        \small
        \begin{tabular}{|p{0.9cm}|p{1.1cm}|p{1cm}|p{1.2cm}|p{1cm}|p{1cm}|} 
            \hline
            \textbf{Assign. No.} & \textbf{\# Comm. (Avg.)} & \textbf{\% Minor} & \textbf{\% Critical} & \textbf{\% Error} & \textbf{\% None} \\
            \hline
            1 & 4.6 & 76.19\% & 21.43\% & 2.38\% & 0\% \\
            \hline
            2 & 5.5 & 70.77\% & 21.54\% & 4.62\% & 3.08\% \\
            \hline
            3 & 9.7 & 74.23\% & 21.65\% & 4.12\% & 0\% \\
            \hline
        \end{tabular}
    \end{adjustbox}
    \caption{\textbf{Survey-Based Classification of ChatGPT's Feedback on Student Submissions.} \textit{Comm.}: Number of Comments; \textit{Minor}: Minor Enhancement Feedback; \textit{Critical}: Critical Improvement Feedback; \textit{Error}: Erroneous Feedback; \textit{None}: None of the aforementioned categories}
    \vspace{-2.1em}
\label{table:code_feedback}    
\end{table}

\noindent In evaluating ChatGPT's feedback on student submissions (Table \ref{table:code_feedback}), a clear trend emerges across three assignments. The bulk of feedback is categorized as "Minor Enhancement Feedback", with percentages of 76.19\%, 70.77\%, and 74.23\% for Assignments 1, 2, and 3, respectively. This indicates ChatGPT's proficiency in suggesting minor improvements to functional code. "Critical Improvement Feedback"  forms about 21\% of the feedback, showcasing ChatGPT's ability to identify significant issues needing major revisions. The "Erroneous Feedback" is minimal, ranging from 2.38\% to 4.62\%, highlighting the model's accuracy in providing relevant feedback. Moreover, the average number of comments per question increases with each assignment, suggesting ChatGPT's capability to offer more detailed feedback for complex coding tasks. This trend demonstrates the model's adaptability to different levels of coding challenges.

\begin{table}[htbp]
\centering
\vspace{0em}
    \begin{tabular}{|p{2.8cm}|p{2.2cm}|p{2.2cm}|}
        \hline
        \textbf{Assignment No.} & \textbf{Effort Score} & \textbf{Modularity}\\
        \hline
        Assignment-1 & \begin{tabular}[c]{@{}l@{}}Average: 0.93\\ Std Dev: 0.36\end{tabular} & \begin{tabular}[c]{@{}l@{}}Average: 1.12\\ Std Dev: 0.85\end{tabular} \\ 
        \hline
        Assignment-2 & \begin{tabular}[c]{@{}l@{}}Average: 1.33\\ Std Dev: 0.28\end{tabular} & \begin{tabular}[c]{@{}l@{}}Average: 1.11\\ Std Dev: 0.32\end{tabular}\\ 
        \hline
        Assignment-3 & \begin{tabular}[c]{@{}l@{}}Average: 0.93\\ Std Dev: 0.46\end{tabular} & \begin{tabular}[c]{@{}l@{}}Average: 1.49\\ Std Dev: 0.86\end{tabular}\\ 
        \hline
    \end{tabular}
    \caption{Ratio = New-Score/Old-Score. The average ratio across each Assignment depicting the increase in scores for each assignment.}
\label{table:code_improvement}
\vspace{0em}
\end{table}

\vspace{-1em}
\subsubsection{\textbf{Objective Assessment:}}
As part of this assessment, we first compute the ratio of the new score ( of modified student code after incorporating ChatGPT's suggestions) and the old score (of original student code) for each question in each assignment. We do this for both metrics for code quality (effort score and modularity). Then for each assignment and for each code quality metric, we compute the average and standard deviation of the ratio of new score and old score across all the questions. The resulting numbers are shown in Table \ref{table:code_improvement}. 

For Assignment-1, the data reveals an average Effort Score ratio of 0.93 (with a standard deviation of 0.36) for the enhanced code compared to the original. This suggests that the complexity of the code remains largely unchanged. However, a notable increase in the average modularity ratio to 1.12 (with a standard deviation of 0.85) indicates that ChatGPT has improved the code's structure, making it more modular and readable while maintaining similar code complexity. 

Moving to Assignment-2, we see an average Effort Score of 1.33 (standard deviation of 0.28), signifying a rise in complexity, likely due to an increase in Logical Lines of Code (LLOC) as ChatGPT enhances the code. Concurrently, the average Modularity score of 1.11 (with a standard deviation of 0.32) reflects ChatGPT's strategy to counterbalance this complexity increase by boosting the code's modularity. 

In Assignment-3, the average Effort Score is 0.93 (with a standard deviation of 0.46), indicating a similar level of complexity to the original code. However, the average Modularity score jumps to 1.49 (with a standard deviation of 0.86), underscoring ChatGPT's continued focus on enhancing code readability and maintainability, while maintaining similar code complexity.

Overall, these findings highlight ChatGPT's approach in code modification: as the complexity of the assignments increases, either due to inherent difficulty or through an increase in LLOC for improvements, ChatGPT compensates by significantly elevating the modularity of the code. This strategy maintains or slightly raises the effort required, however it also ensures that the code quality is enhanced, making it more structured, readable, and maintainable.

In terms of code correctness, we observed minimal changes in the scores. This lack of significant improvement matches with our subjective assessment wherein we found that most of the suggestions given by ChatGPT aimed at minor enhancements in the code. This led to increased code quality but no observable improvement in the code correctness.  
\section{Limitations}
This study is subject to certain limitations that merit consideration. In some experiments, we were limited by a relatively small sample size that may not be fully representative of broader classroom settings. Additionally, we encountered occasional inconsistencies in the responses from ChatGPT-3.5. Due to limited resources, we had to rely on this version of the model despite its occasional unpredictability. Substantial prompt engineering is still required to optimize its performance and ensure response consistency. Additionally, the feedback offered by ChatGPT lacks the level of personalization typically provided by human teaching assistants. This is primarily due to the fact that ChatGPT's prompts were not informed by any historical data on individual students' coding habits or areas of weakness. Lastly, the inherent variability in ChatGPT's responses means that achieving identical results in repeated tests can be challenging. As a result of these factors, we believe that further research may be required to ascertain if the findings of our study can generalize to other classroom environments.

\section{Conclusion and future work}

 An introductory programming course is a common course in many colleges which often has large enrollments. As the course generally involves many programming assignments, for evaluating assignments and giving feedback to students, often a large number of TAs are needed, which is a significant challenge and overhead which brings in issues of consistency in evaluation. Clearly, teaching an introductory programming course can benefit tremendously if the TA functions can be partially automated and the quality of TA support is improved. 

In this paper, we explored the potential of using ChatGPT,  which is expected to significantly impact education, as a virtual Teaching Assistant (TA) in an Introductory Programming Course. We evaluate ChatGPT's capabilities for two essential TA functions: grading student code submissions and providing feedback to undergraduate students in an introductory programming course.  We evaluate these capabilities of ChatGPT through two experiments, involving three homework programming assignments from the CS1 course at our institution, where Python is the primary language. The course had approximately 650 students, with each assignment comprising various programming tasks of differing complexities.


In the first experiment, we assessed ChatGPT's proficiency in grading student code submissions for both code correctness and quality. The code correctness scores provided by ChatGPT were compared with those assigned by human TAs. Similarly, for quality assessment, we compared ChatGPT's scores with those generated by the Radon package. Our analysis indicates that, at present, ChatGPT is not completely reliable in evaluating either the correctness or the quality of code.

In the second experiment, we evaluate the potential of ChatGPT to provide suggestions to students for code improvement. The suggestions provided by ChatGPT were evaluated by four junior and senior year students who were previously TAs and high performers. The data shows that ChatGPT provides many valuable suggestions leading to increased modularity and improved code structure. 

It is widely recognized that ChatGPT has the propensity to produce distinct responses when presented with identical questions framed using differing prompts, even if the prompts vary only slightly \cite{Reeves2023Parsons}. In our future work, we intend to conduct further experiments that specifically explore the impact of prompt engineering on the performance of ChatGPT in the context of grading student code and giving feedback to students. Ultimately, our goal is to create a virtual teaching assistant (bot-TA) powered by large language models (LLMs) like ChatGPT, which can pinpoint individual student's areas of difficulty and provide personalized tutorials and reference materials tailored to their needs. Moreover, this bot-TA could facilitate more impartial evaluations by eliminating the potential bias that arises when multiple human teaching assistants are involved in assessing students' work. 


\bibliographystyle{ACM-Reference-Format}
\bibliography{chatgpt-ref}

\appendix

\end{document}